\documentclass[11pt,a4paper,twocolumn]{article}
\usepackage{mathtools}
\usepackage[left=2.25cm,right=2.25cm,top=2.5cm,bottom=3cm]{geometry} 
\usepackage{hyperref}

\title{\textbf{Causes of discomfort in stereoscopic content: a review}}
\author{\textbf{Kasim Terzi\'c and Miles Hansard}\\[1ex]
School of Electronic Engineering \& Computer Science\\[.5ex]
Queen Mary University of London}
\date{August 2016}

\begin{document}

\maketitle

\begin{abstract}
This paper reviews the causes of discomfort in viewing stereoscopic content. These include objective factors, such as misaligned images, as well as subjective factors, such as excessive disparity. Different approaches to the measurement of visual discomfort are also reviewed, in relation to the underlying physiological and psychophysical processes. The importance of understanding these issues, in the context of new display technologies, is emphasized.
\end{abstract}

\section{Introduction}
The availability of stereoscopic displays and content has increased greatly, during the last ten years.
In addition to the many new films produced in 3D, there is a market for re-releases of older movies adapted to 3D, and a strong
push by TV manufacturers toward stereoscopic viewing. There are indications, however, that there has been a decline
in popularity of 3D movies in the last two to three years, and many attribute this decline to complaints of
discomfort associated with viewing 3D content.

A 2013 study on 433 subjects estimated that 14\% of people suffer from some discomfort symptoms, 
with additional 8\% reporting discomfort related to wearing the special equipment needed, such as 3D glasses.
Symptoms were reported to be mainly headache and eyestrain \cite{read13user}.
Another study from the same year on 524 subjects reported that more than half of the viewers 
suffer from some symptoms \cite{solimini13plos}. 
It seems plausible that this discomfort may play a role in the declining popularity of stereoscopic films.

Results like these (obtained on smaller samples) were known
for many years \cite{kooi04visual} and have led to research on the causes of discomfort. This research has flourished 
in the past few years, with a number of papers which summarise the main findings 
\cite{ukai08visual,lambooij09visual,howarth11potential,tam12visual,bando12visual,urvoy13how,kondoz15novel}. 
A number of guidelines
have appeared for cinematographers \cite{mendiburu09,zilly11production}, and quality assessment 
of stereo images and videos became important, though typically with a stronger focus on
image quality than viewer discomfort \cite{winkler13stereo}. 

This report presents a new and comprehensive overview of the findings in this field. We feel that such a review
is necessary for several reasons. First, it is a very active field, and many new facts have surfaced in recent 
years. More than half of the results presented here are from the last five years, and a third were published since 
2013. These recent advances are not covered by previous reviews. 
Second, there has been a recent push into obtaining physiological insight by using objective
measurement techniques such as brain scans. 
We feel that this important part of the equation has not been adequately addressed 
by previous reviews of discomfort literature. A recent review on objective measurements dealt with effects
of stereoscopic and 3D viewing in general and did not specifically focus on comfort or 
fatigue \cite{park15overview}. To our knowledge, this is the first work to attempt
to include this data into a wider discussion of discomfort. Finally, there is a strong push toward new,
non-traditional displays, such as head-mounted displays (HMDs) for virtual reality, mobile devices and smart
glasses \cite{shibata13comfortable}. Affordable consumer HMDs are set to arrive in 2016, and the immersive
nature of VR applications could lead to strong visual fatigue. We discuss specific issues related to these new
technologies.

This paper is structured as follows. In Sec.~\ref{secVisual} we give a brief overview of the visual 
system, and discuss how stereoscopic viewing can stress it in unusual ways. In Sec.~\ref{secMeasure} we
discuss subjective and objective measurement techniques used to assess visual discomfort. In Sec.~\ref{secContent}
we summarise the type of content strongly associated with discomfort, most of which was obtained through numerous
subjective studies. In Sec.~\ref{secScans} we summarise the current understanding of
the physiological basis of discomfort, based on objective measurements. 
Sec.~\ref{secHMD} discusses considerations specific to emerging
technologies such as HMDs, followed by a discussion in Sec.~\ref{secDiscussion} and a conclusion.

\section{Visual system and stereoscopy}
\label{secVisual}
Our visual system evolved to process natural scenes. In this section, we briefly introduce the human
visual system and point out how stereoscopic video differs from natural viewing and can thus 
provoke an unnatural response. Then in later sections, content related to 
discomfort and measurements of physiological factors will be analysed in more detail.

\subsection{Eye movements and geometry}
The basis of stereo vision is the ability of our visual system to fuse the left and right views into a 
single cyclopean view of the scene. The apparent displacement of objects when viewed from two 
positions is called \textit{parallax} and it gives rise to \textit{retinal disparity}, the difference in
location of the object's projections on the left and right retina.

Opposite eye rotations are called vergence, and one often speaks of convergence (eyes rotating
inward) or divergence (eyes rotating outward). 
In natural viewing, the eyes converge on an object of interest, so that the object's 
retinal projections have near-zero retinal disparity. The locus of zero disparity is called the horopter
Within a small region around the horopter, where disparity is small, the visual system perceives a single
object. This area is called Panum's fusional area 
\cite{banks12stereoscopy},
and measurements suggest that disparities of up to 0.5 degrees can be fused \cite{yeh90limits}, though
the specific value depends on the location on the retina and is closer to 
0.1 deg around the fovea \cite{lambooij09visual}.
In practice, hard limits are difficult to measure because disparity sensitivity is tied to luminance
and contrast \cite{didyk12luminance}, as well as spatial frequency content \cite{schor84binocular}.

In stereoscopic viewing, other types of eye movement may result from incorrect camera geometry. For example,
vertical offset between the left and right views causes vertical vergence \cite{howard00effects}. This is 
discussed in more detail in Sec.~\ref{secContent}.

\subsection{Accommodation and vergence}

The process of vergence is related to accommodation, the process by which the eyes bring the
converged object into sharp focus. 
Accommodation error must be within $\pm$0.25D (diopters) for the object to appear sharp \cite{hoffman08vergence};
at the same time, accommodation results in blurring of objects which are difficult to fuse. 
Accommodation and vergence are known to be tightly coupled 
\cite{fincham57reciprocal,martens59observations,schor99influence} 
and can be modelled by a dual-parallel feedback-control system \cite{lambooij09visual}, the result of which
is that both vergence and accommodation are faster when coupled \cite{cumming86disparity}.
Moreover, vergence speed is dependent on initial position \cite{alvarez05divergence}, so small depth 
adjustments are faster.

The zone in which an object is both sharp and has low disparity is termed the Zone of Clear Single 
Binocular Vision (ZCSBV) \cite{fry39further}. It measures maximal decoupling while maintaining a clear, 
single, binocular percept. ZCSBV does not guarantee comfortable viewing. 
Percival's zone of comfort \cite{percival92relation} is defined as the middle third of ZCSBV, and
the alternative Sheard's zone \cite{sheard34prescription} is a bit better predictor for 
exophores \cite{sheedy77phoria}. These measures were developed for natural viewing, and correlate
well with stereoscopic viewing for near distances, but not for far distances \cite{shibata11zone}.

Stereo 3D breaks the coupling between vergence and accommodation because the eyes keep a sharp
focus on the screen depth, while vergence is varied to process varying disparities.
Shibata's Zone of Comfort \cite{shibata11zone} was specifically developed to address this issue.
Even within the comfortable zone, problems can arise if there is much variation in screen 
disparity \cite{nojiri03measurement,nojiri06parallax}. 
More discomfort is felt when the conflict changes rapidly \cite{banks13insight,kim14rate}.
 
The effects of the vergence-accommodation conflict are numerous. 
An experiment by Hoffman et al.\ showed that it significantly affects discomfort 
and degrades depth perception \cite{hoffman08vergence}.
It increases the time required for fusion \cite{watt05focus,akeley04stereo,kim12effect}, which affects fast-moving
objects and cuts between scenes. After an hour of watching stereoscopic material, 
there is a measurable decline in accommodation response suggesting fatigue \cite{yano02study}.
Accomodation response is normal if the object in focus is within the depth of field \cite{hiruma93accomodation},
otherwise it seeks a more comfortable and less stressful state \cite{okada06target}.

This topic is subject to active research. 
A recent study \cite{shiomi13simultaneous} indicates that the conflict occurs only 
with near displays and is not a factor for many viewing scenarios. 
Another study found the effect to be worse on small displays \cite{cho12measurement}, which agrees with this
finding.

\subsection{Depth cues}
The visual system uses a number of different cues to determine depth. They include
blur, shading, perspective, disparity, haze and motion \cite{palmer99}.
In addition to these bottom-up cues, top-down effects also influence depth 
perception \cite{buelthoff98topdown,welchman05three}. While
ambient illumination is not important for depth perception \cite{polonena11effect},
high-contrast lighting helps to enhance the apparent depth of a scene \cite{banks12stereoscopy}.

There is no unified model of how the depth cues are combined \cite{howard02depth}
but the process is often modelled in the Maximum Likelihood framework \cite{fisher25theory}.
The visual system can resolve conflicting cues \cite{stelmach00stereo}, but when there is strong conflict, 
rivalry dominates \cite{heuer87apparent}. 
In terms of discomfort, it has been shown that cue combination using Minkowski
summation is a good predictor for overall levels of visual discomfort. The overall level of perceived
discomfort is determined by the most significant discomfort factor in a winner-takes-all manner
\cite{lee14experimental}.

Disparity is one of the most important cues, and forms the basis of stereoscopic vision.  
Discrimination thresholds are higher for larger corrugations \cite{howard02depth} and 
larger disparities \cite{blakemore70}. Disparity sensitivity is similar to the 
contrast sensitivity function \cite{bradshaw99sensitivity}.
Depth is dominated by distribution of disparity contrasts, strong at discontinuities and weaker
at ramps \cite{brookes89}, which might explain the apparent ``flatness'' of stereoscopic 3D.

Stereoscopic 3D can provide inconsistent depth cues. In natural viewing, focus blur is an important
cue, but it is absent in stereoscopic 3D. This also results in a lack of blur gradient along ramps
and smooth depth transitions, which adds to the perceived flatness of the scene \cite{watt05focus}. 
The parallax due to head movement is completely absent, though on-screen motion still provides a strong dynamic
parallax. Other incorrect cues include fixed accommodation, and wrong sizes of observed objects, leading to 
the ``puppet theatre effect''.

\subsection{Visual cortex}
The primary visual cortex V1 receives retinal images via the Lateral Geniculate Nucleus and processes them
trough a combination of simple, complex and end-stopped cells, for which efficient computational models
exist \cite{terzic14realtime}. Even at this early stage, the left and right stimuli are processed
together and cells associated with the same retinal position in left and right views are located close to
each other in the cortex, hinting at stereo disparity processing early in visual cortex \cite{neri04stereoscopic}.
From here, coarse disparity is associated with the ``dorsal'' pathway responsible for 
localisation and spatial layout, via cortical areas V2, V3 and MT \cite{parker07binocular}, while
fine disparities aid shape and object recognition in the ``ventral'' pathway, via V2, V4, and IT.
Numerous computational models for cortical disparity calculation have been proposed, including 
phase-based stereo, disparity energy \cite{terzic13biological}, and sparse matching of end-stopped cell 
responses \cite{terzic14realtime}, but disparity processing is still subject to intensive research.

Top-down influence on depth perception has long been established 
\cite{buelthoff98topdown,welchman05three}, which hints at the involvement of higher cognitive processes
in depth perception. Mental fatigue in trying to resolve conflicting cues
is a possible cause of discomfort.

\subsection{Eye fixations and attention}
Our visual system processes the scene sequentially through a sequence of saccades, preferring ``salient''
parts of the image. Depth is known to strongly affect the salience of 
image regions \cite{ma15learningbased}, leading to a number of salience measures which incorporate
depth \cite{quan11importance,wang13computational}. 
There is a correlation between salience and discomfort \cite{cho12subjective,jiang15depth}, and computational
models of discomfort perform better once salience is taken into 
account \cite{sohn13disparity,jung13predicting,wang13visual}.

\section{Techniques for measuring discomfort}
\label{secMeasure}

In order to tackle discomfort, we must first be able to measure it. While there is a good review
of measurement techniques related to stereoscopic vision in general by Park and Mun \cite{park15overview},
it does not focus on discomfort specifically. Similarly, literature reviews which focus on Quality Assessment
of stereoscopic video are primarily concerned with the perceived quality of the video and viewing
preference, of which discomfort is only a factor \cite{moorthy13survey,su15visual}. In this chapter, 
we quickly summarise the types of measurements used to assess visual discomfort before we discuss the
major findings in the following sections.

There are two major types of measurement: subjective measurement, which involves asking the viewers to
assess the amount of discomfort by filling out questionnaires or moving sliders; and objective measurements, 
which observe the body's response to stereoscopic video through eye trackers and brain scans. 
Subjective measurement is crucial for determining which content and viewing conditions cause discomfort and
detecting that discomfort is present. Objective measurement can help us understand the underlying physical 
processes which lead to it by comparing the responses during normal viewing and uncomfortable viewing. 
Lambooij stresses the difference between discomfort, which is subjective, and fatigue, which is objective
and measurable \cite{lambooij09visual}.

Measurement is difficult because, people are more likely to disagree about quality of depth
than the quality of flat images \cite{chen12image}, so results tend to be less consistent. Discomfort also
depends on stereoacuity (more discomfort for better stereoacuity) \cite{kim13influence}. It is not clear
how discomfort is affected by age: some studies found no big difference between children and 
adults \cite{polonen13stereoscopic,zeri15visual} but other studies suggest that this only holds for 
large disparities and medium ambient illumination \cite{wang15relationship}. Yang et al.\ found
that younger people are disproportionately affected \cite{yang12stereoscopic,yang11individual}.
Women are found to be more strongly affected  than men \cite{zeri15visual,yang11individual}. Also, a strong
hereditary influence has been suggested in a recent study \cite{lee12individual}.
All this suggests that more personalised media will be necessary in the future.

\subsection{Subjective measurement}
The only certain way to know if someone is comfortable is to ask them. Subjective 
measurement also has the benefit of easily obtaining many sample points (some studies used hundreds
of subjects). The problems involve the inconsistency (questionnaires are subjective by their very
nature) and fatigue associated with long tests.
Subjective evaluation is important because it allows us to identify problematic content, and most insights
presented in the next section were obtained from user studies.
A number of protocols have been applied to discomfort measurement, such as
the Binocular Just Noticeable Difference model to calculate distortion visibility 
threshold \cite{hachicha13stereo}. 

The most important method applied to discomfort measurement is probably the
ITU recommendation BT.500-10, which measures a wide range of image impairments on a scale from
``imperceptible'' to ``very annoying'' \cite{itubt500}. 
For example, it has been applied for measuring the effect of crosstalk \cite{barkowsky13towards}, 
but it was originally designed to measure image quality, not comfort.
Even when it is modified by researchers, this recommendation is still important because it defines many 
test conditions which can help to improve consistency between tests.
The ATSC suggested using a single Likert scale ranging from ``very comfortable'' to ``very uncomfortable'' 
\cite{atsc}.
Hoffman et al.\ used a combination of five-point Likert scales covering aspects such as how tired the eyes
feel and how clear the vision \cite{hoffman08vergence}.
The Convergence Insufficiency Symptom Survey (CISS) by Lambooij et al.\
addressed different aspects of discomfort, such as double vision and and sleepiness \cite{lambooij09visual},
and Yang et al.\ added psychological factors such as impaired memory, disorientation, dizziness and vertigo 
\cite{yang12individual}. The Stereoscopic Discomfort Scale (SDS) of Bracco et al.\ combines previous
measures and extends them with new ones in order to create a more complete standard \cite{bracco13investigating}.
Unfortunately, none of these scales has been widely adopted, and many are still based on the ITU scale, making
comparisons difficult.

Questionnaires may be completed after a video clip has been viewed. In order to obtain near real-time
measurements, researchers have turned to continuous response measurement techniques \cite{biocca94continuous}.
ITU BT.500 includes the Single Stimulus Continuous Quality Evaluation (SSCQE) protocol first
introduced by Hamberg and De Ridder \cite{hamberg95continuous} which has been applied to stereo discomfort 
measurements, e.g. \cite{lambooij11visual,yano02study,ijsselsteijn98perceived}.
In Quality Assessment, a number of metrics have been developed to predict discomfort
 \cite{voronov13methodology}

\subsection{Objective measurement}
Objective measurements have the advantage that they can be automated, and tell us more about physiological
causes of discomfort. However, they must correlate well with subjective results in order to be useful,
and this is often difficult.

A wide variety of physiological measurement techniques exist, but only few have successfully applied to
predicting discomfort. It is known that ECG can measure cognitive load \cite{haapalainen10psycho}, which
is one measure of strain and fatigue. Brain responses can be measured via EEG and fMRI scans, and both
have been shown to correlate with visual fatigue. Finally, ophthalmological measurements such as eye
movement, pupil size and vergence have been used to measure these factors directly and determine their
correlation with visual discomfort. 
A summary of related findings is given in Sec.~\ref{secScans}.

\section{Content associated with discomfort}
\label{secContent}

It has been suggested that visual discomfort is caused by the instability of 
the perceived world \cite{hwang14instability}.
Stereo 3D is an imperfect simulation of the real world, and as discussed in Section \ref{secVisual},
this can cause unnatural strain on different parts of our visual system. Therefore, it follows that
content has a large influence on perceived discomfort: the type of content which forces the visual system
to act in an unnatural way is more likely to cause discomfort.
It is important to note here that discomfort and perceived image quality are not the same thing. While there is
much research on quality assessment of stereoscopic images and video, the perceived image quality is not 
necessarily a guarantee that extended viewing will be comfortable.

A number of different surveys performed over the years have identified the types of content most likely
to cause discomfort. They are 
incorrect viewing geometry \cite{bereby99perception},
vertical disparity \cite{kooi04visual}, 
excessive horizontal disparity and rapidly moving objects \cite{yano02study,lambooij11visual},
crosstalk \cite{kooi04visual,wang10cross},
unnatural blur \cite{okada06target,kooi04visual}, window violations \cite{lopez13stereoscopic}, 
fast motion in depth \cite{peli99,lambooij11visual,lambooij13impact}, and image distortion from incorrect 
pre-processing \cite{woods93image}.
In the following, we discuss these factors in more detail and attempt to quantify them.

\subsection{Incorrect viewing geometry}

The human visual system evolved to view natural scenes, and is optimised for this particular constrained
viewing geometry. When artificially created stereoscopic images are presented to the eyes, these constraints
are violated, leading to additional stress on the visual system
\cite{bereby99perception}.
One of the first analyses of image distortion in stereo viewing was given by Woods \cite{woods93image}.

In natural viewing, the distance between the eyes (interaxial distance) is fixed, but the distance between 
two cameras in a stereo configuration can vary.
Cinematographers often modify the separation of the cameras for each scene 
separately to adjust the amount of disparity in a shot \cite{holliman06cosmic,murray06three}. 
Surprisingly, viewers do not seem to be very sensitive to this; they ignore motion and stereo cues 
in favour of a fictional stable world~\cite{glennerster06humans}.

The primary depth cue in stereoscopic content is horizontal disparity -- the horizontal
offset of an object in one view compared to the other. The fixed position of the eyes ensures that horizontal
disparity dominates regardless of the position of the head, and our visual system is particularly good
at processing horizontal disparities. When two cameras in a stereo configuration are misaligned,
the left and right images are no longer vertically aligned and this has been identified as a major
factor in discomfort \cite{kooi04visual}. The eyes adjust to this situation through
vertical vergence where the two eyes rotate vertically in opposite directions \cite{allison07analysis}. 
This movement is not natural and eventually leads to fatigue \cite{howard00effects}. 
Acceptable levels of vertical disparity are considered to be about 15 arcmin, corresponding to a torsional 
disparity of about 30 deg of relative orientation \cite{tyler12three}.
Displays which show left and right images on alternating
rows provoke the same response, but this effect is considered too slight to 
cause eyestrain \cite{banks12stereoscopy}. 
Vertical misalignment can be easily fixed during acquisition, but it also surfaces with ``good'' videos.
If the head is not kept vertical, a rotation of the two views will result on most displays \cite{kane12visual}.
This means that simply tilting the head in natural viewing can lead to discomfort.

A more complex situation occurs when the two images are not simply offset, but shot from a different
perspective, as in an toe-in configuration. This also causes 
vertical disparities \cite{mcallister93,woods93image,kooi04visual,ware98dynamic}, 
but causes additional
problems. 
Such content can also be visually confusing, because of implicit cues about camera alignment \cite{longuet81computer,ogle38induced}. 
Toe-in camera configuration gained popularity in part because it was thought that it reduces the need for cropping.
Material captured by cameras in a parallel configuration will inevitably have parts in each image which are 
not visible in the other.
If an object appears in these regions, this leads to window violations \cite{lopez13stereoscopic}, which
are a major cause of discomfort \cite{zhou14comfort}. 
Because of these factors, toe-in filming is discouraged today in favour of parallel cameras followed by cropping.
\cite{banks12stereoscopy}. 

Incorrect viewing geometry puts extreme stress on the visual system. Luckily, many of the worst aspects can be
eliminated if well-calibrated and properly aligned cameras are used during acquisition. The problem with 
head rotation during viewing is, unfortunately, much more difficult to solve without HMDs.

\subsection{Crosstalk}
Crosstalk (or `leakage') is the process by which one image is combined with another during playback. 
The resulting effect, where objects are seen in double, is called ``ghosting''.
Huang distinguishes between system crosstalk (related to the device) and viewing crosstalk (related
to the content) \cite{huang06crosstalk}. This effect is entirely unnatural and completely caused
by imperfect technology and as such, it is reported as one of the most annoying factors in stereo 3D
\cite{kooi04visual,wang10cross}.

The wide range of causes of crosstalk means that there is no unified solution.
It depends on the specific display, specific shutter glasses (if used) and the viewing 
angle \cite{zafar14characterization} and due to 
the wide range of available display equipment, reported results are not always consistent.
Passive glasses are traditionally considered more prone to crosstalk (especially colour filter-based anaglyph
glasses), and there are systems which
claim that shutter glasses eliminate ghosting completely by some measures \cite{chen11system}.
On the other head, a study from 2013
claims that crosstalk is lower on passive displays than on active displays \cite{yun13evaluation}.
Yet another study found no major difference between active and passive stereo \cite{wang11crosstalk}. It is widely
accepted, however, that crosstalk contributes to discomfort and can be removed by technological means.

It has been claimed that around 20\% crosstalk is considered acceptable with mirror-type 
displays \cite{wang13computational}, but this seems high for normal stereo content.
One study finds that 15\% is considered annoying \cite{seuntiens05perceptual}, while another one
recommends less than 10\% \cite{nojiri04visual}. Quality impairment is sufficient to affect
depth perception with as little as
4\% \cite{tsirlin11effect}, but there is no proof that such levels of crosstalk cause discomfort.
Annoyance due to crosstalk increases with increasing disparity \cite{wang11crosstalk}, 
increasing camera base distance \cite{seuntiens05perceptual,xing12assessment}
contrast \cite{wang11crosstalk}, and scene content \cite{xing12assessment}.

Crosstalk negatively affects depth perception 
\cite{tsirlin11effect,tsirlin12crosstalk} which can cause additional strain.
Perceptual crosstalk tests show that interocular crosstalk is a function of spatial frequency
\cite{ijsselsteijn05human}.

Where present, crosstalk can be masked by perceived motion blur, especially at low binocular parallax, 
which limits the crosstalk-induced image quality degradation \cite{wang12effect}.
It has also been argued that the blurring caused by crosstalk can reduce the vergence-acommodation 
conflict so small amounts of crosstalk can be beneficial in practice \cite{lambooij09visual}, but
there are more effective depth-of-field methods for dealing with this problem.

\subsection{Excessive Disparity}
\label{secExcessive}

Unlike vertical disparity, our visual system is well-equipped to deal with large horizontal
disparities. But even here, there are limits to what the visual system is capable of fusing,
and the failure to fuse can be very uncomfortable if it persists over long periods of time.
It is easy to perform acquisition in a way which results in excessive disparities, so care is needed.
Experiments have shown that 
there is a comfortable viewing range \cite{wopking95viewing,jones01controlling} which 
limits the allowed horizontal disparities.
In addition to absolute disparity, the relation to object is also important, as smaller 
stimulus width causes more strain \cite{lee13effect}, as does a large disparity between
the foreground and the background \cite{kim12effect}.

In cinematography, there are guidelines. Mendiburu cites the 3\% rule 
\cite{mendiburu09}.
Lambooij argues for one degree of screen disparity \cite{lambooij09visual}.
Williams gives the maximum disparity as 25\% (in front) or 60\% (behind) 
of the viewing distance \cite{williams90new}.
Excessive disparities are closely tied to the accomodation-vergence conflict. Shibata et al.\ 
determined that the comfortable limits are 2-3\% of the screen width for crossed (in front) 
and 1-2\% for uncrossed (behind) disparities \cite{shibata11zone}. While differently stated, all 
of these measures are quite comparable and serve to illustrate that the range of depths
in stereoscopic 3D must be tightly controlled to a sub-volume centred around the screen
depth. This severely limits the range of disparities allowed for a comfortable viewing
experience.

The difficulty of keeping disparities within this range during acquisition has spawned a number
of computer algorithms capable of automatically adjusting the disparity range of existing
stereo content \cite{lang10nonlinear,yan13depth,qi13shift,jung14visual,sohn14local,oh15visual}.

\subsection{Blur}

Blur plays an important role in depth perception \cite{zhang15visual}, and it is also crucial in
reducing discomfort. 
Unnatural blur is often cited as a cause of viewing discomfort in stereo images \cite{okada06target,kooi04visual}.
However, it has been recently argued that artificial blur itself does not induce discomfort when applied to 
a scene, so it is the inconsistency with the acommodation process that is the likely cause \cite{ohare13visual}.
Wopking proposed that depth of field helps discomfort \cite{wopking95viewing}, which was later
proved by Blohm \cite{blohm97stereoscopic}, who showed that test subjects prefer images where only a subvolume
corresponding to a limited range of depths is in sharp focus. The rest of the scene is
blurred, and corresponding disparities masked, resulting in a limited disparity range as described
in the previous section.

In natural viewing, the visual system keeps the object of interest in sharp focus through the 
process of accommodation to this specific distance. Since the vergence and accommodation mechanisms
are coupled (see Sec.~\ref{secVisual}), the object in focus should have near-zero disparity. 
Objects which are in front or behind this plane are blurred. The benefit of this process is that 
objects exhibiting large disparities are blurred more strongly, and our visual system does not attempt
to fuse them. Thus it is the absence of accurate blur in most stereoscopic video that contributes
to discomfort by overloading the visual system. 

An additional problem is caused by the absence of blur gradient along depth gradients. 
Disparity as a cue is strongest at sharp boundaries, and the absence of blur-based cues results in 
an impression that all objects in the scene are flattened. The visual system works hard to try to
resolve the conflict with the high-level expectation, which can lead to fatigue.

Blur is a difficult aspect to address. Since stereoscopic video is presented at a fixed distance,
the eyes will naturally accommodate to this distance, thus losing blur as an important depth cue. Systems
based on eye tracking \cite{duchowski14reducing} and selective blurring \cite{leroy12visual} have either 
failed to improve viewing comfort, or have had to sacrifice image quality.

\subsection{Motion in depth}
It is not clear that movement in stereoscopic films is uncomfortable \textit{per se} 
\cite{li14motion}, but there are particular types of movement associated with discomfort. 
For example, Yano et al.\ report that in-plane motion within the zone of comfort does not
lead to more fatigue than 2D viewing \cite{yano04two}.

Most authors single out motion in depth as particularly uncomfortable,
as first studied in detail by Speranza \cite{speranza06effect}. There seems to be complete agreement among
researchers on this point \cite{peli99,lambooij11visual,lambooij13impact,lopez13stereoscopic}. 
The worst culprit is motion between positive and negative disparity \cite{speranza06effect,lopez13stereoscopic}.
According to much research, slow motion in depth is more comfortable than fast motion in depth, 
which should be avoided \cite{peli99,lambooij11visual,lambooij13impact}. 
But the evidence here does not seem to be conclusive. 
Recent research by Hartle et al.\ examined viewer preferences for different types of camera movement
and found that there is some preference for faster movement \cite{hartle15preference}, specifically for the
movement-in-depth case. They conclude that viewer preferences are complex and do not necessarily
exhibit direct relation with an individual cause.

In natural viewing, an observer will analyse the scene by sequentially focussing on different scene
objects, which requires occulomotor adaptation to corresponding depths.
The speed of vergence depends on the disparity jump \cite{rashbass61disjunctive}, but
the coupling between vergence and accommodation is optimised for natural viewing.
From the earlier
discussion of the vergence-accommodation conflict, it was seen that the eyes' adjustment to depth
is slower for stereo 3D than in natural viewing 
\cite{cumming86disparity,watt05focus,akeley04stereo,hoffman08vergence}, which affects any change in observed
depth. 
Recent results provide evidence that dividing attention between multiple salient objects is a cause of 
discomfort \cite{zhang15visual}. This has also been shown for movement consisting of steps in depth
\cite{yano04two} and depth jumps \cite{lopez13stereoscopic}. This effect severely constrains the make-up
of stereoscopic video. In addition to minimising the depth range in a scene, which was discussed in 
Sec.~\ref{secExcessive}, it also means that sharp cuts are a potential cause of discomfort and that
content creators should ensure that such cuts do not result in sharp changes in disparity.
A qualitative study of combinations of factors found that frequency and abruptness of disparity 
change were the strongest cause of discomfort \cite{kim13qualitative}.

Other types of motion can also lead to problems. It has been noted that rapidly moving objects 
have an effect on viewing comfort \cite{lambooij11visual}, and research showed that 
relative disparity \cite{li11quality} and velocity \cite{lee11visual} are main factors for visual discomfort 
in the case of planar motion. Similar findings were reported by Tam \cite{tam12visual} and by Du, who proposed
a comfort metric which incorporated 3D motion \cite{du13metric}. All of this suggests that stereoscopic
videos should be more constrained, not only in depth range, but also in the speed of movement. The popularity
of fast action movies with quick cuts, many of which are shown in 3D, seems to present a problem for
comfortable viewing.

\subsection{Visual tolerance}
After outlining all the content which causes or exacerbates viewing discomfort, we are happy to report
that there is also content which is not problematic. A 2015 study on 854 subjects
showed that seating position (in a cinema) did not matter. It also found that more recent films caused
less discomfort, suggesting that acquisition is improving in line with the findings and best
practices outlined in this chapter \cite{zeri15visual}.

While difference in zoom between the left and right views can cause discomfort
by introducing vertical disparities and create 
the appearance that the scene is slanted and cause problems \cite{ogle38induced},
it was found that difference in spatial resolution is not crucial \cite{stelmach00stereo},
and neither are differences in interocular luminance \cite{boydstun09stereoscopic}. Since
this is common in natural viewing (e.g. with people who are near-sighted on one eye only), our visual system
may have evolved to deal with such situations.

\section{Physiological factors of discomfort}
\label{secScans}

It has been suggested that ECG measurements can indicate an overload of the autonomic nervous system
\cite{park15overview}, and a recent study found a correlation between ECG readings and visual discomfort
caused by stereoscopic viewing \cite{kim13autonomic}. Not all studies found such a correlation, for example
no difference was found in ECG LF/HF ratio \cite{naqvi13does}.
Most of the objective measurements have concentrated on eye responses and brain scans.

\subsection{Ophthalmological factors}

It has been argued that oculomotor factors are predominant in visual symptoms and there is some correlation 
between discomfort and microsaccadic movements \cite{vienne12visual}. 
Blinking is correlated with eyestrain \cite{lee10comparative,kim11stereoscopic}, 
but it is not always a good predictor of discomfort \cite{li13visual}. Researchers have, however
succeeded in mapping eye blinks to subjective discomfort scores \cite{cho12subjective}.
Another study using EOG \cite{yu12eog} detected more saccades and blinks for stereoscopic 3D than for 2D 
material, which could be one of the causes of fatigue.
More recently, several studies showed that a statistical analysis of eye tracking data 
can predict discomfort \cite{iatsun15investigation,cho15analysis}.
Kim and Lee found that visual attention is strongly
affected by visual fatigue and  
they incorporated this insight into their Transition of Visual Attention model \cite{kim15transition}.

Care is needed, because while the link between blinking duration and number of saccades with visual fatigue 
has been established,
measurements of the pupil diameter and fixations are not always precise enough and are highly dependent on 
content \cite{iatsun12investigation}. It is likely that technological improvement will resolve these 
problems \cite{park15overview}.
It has been suggested that reading speed can be a useful proxy for fatigue, as it decreases 
as a result of visual fatigue \cite{lambooij12assessing}.

Additional information has been obtained through direct measurements of vergence and accommodation \cite{neveu12vergence}.
It has been observed by Ukai and Kato that vergence and accommodation are impaired when
observing stereoscopic 3D \cite{ukai02use}. After an initial adjustment of accommodation to correspond
with the change in vergence, accommodation returns to the screen surface, causing an oscillatory
behaviour in both accommodation and vergence. Cho et al. showed that accommodation depends on the
amount of blur \cite{cho96study}, and subsequent research confirmed that low-pass filtering influences
accommodation response \cite{torii08dynamic}. Therefore, selective blurring may help reduce discomfort. 
Several studies found that the vergence-accommodation mechanism is impaired 
after prolonged viewing of stereoscopic material \cite{schor87fatigue,saito94physio}, in particular the 
natural accomodative response is slowed down \cite{iwasaki02effects,suzuki04effects}.

Kim et al. directly measured fusion time and found that discomfort depends on the parallax difference 
between foreground and background \cite{kim12effect}. A wide range of ophthalmological measurements
by Wee et al. included
 near point of accommodation (NPA) and convergence (NPC), amplitude of fusional convergence and divergence,
 tear break-up time and temperature of ocular surface, and angle of phoric deviation
 \cite{wee12ophtalmological}. 
 They found that accommodation and binocular vergence are the predominant factors of discomfort.

\subsection{Neural factors}

While there have been recorded attempts of using different types of technologies including 
MEG \cite{hagura06study}, most research has concentrated on real-time (but less precise) EEG measurements
and more detailed (but slower) MRI imaging techniques.

\subsubsection{EEG}

Cortical measurements of 3D-induced visual fatigue date back at least to 
Yamazaki et al. \cite{yamazaki90quantitative}, who found P100 latency to be a good
predictor of fatigue. P100 relates to event-related potentials (ERP) and refers to a cortical response
to an event after an approximate 100ms delay. 
The P100 latency was shown to increase in visual evoked cortical potentials in
the left (LO), right (RO) and middle (MO) parts of the occipital lobe, and the vertex (Cz)
\cite{emoto05repeated}. These effects disappeared after a rest.
Similarly, an increase in P300 and P700 was also observed after prolonged stereoscopic
 viewing \cite{li08measurement}.
Significant reduction in P600 potentials and increase in P600 latencies was observed by Mun et al.\ 
\cite{mun12ssvep}. They additionally measured steady-state visually evoked potentials (SSVEP), which
are more commonly related to low-level processing. Significant reduction in attend/ignore ratios
was obtained after stereoscopic viewing in the parietal area P4 and occipital area O2.
An uncomfortable stereoscopy correlates with a weaker negative component and a delayed positive component in ERP
\cite{frey14assessing}. 
Interestingly, passive polarised displays do not seem to affect ERP \cite{amin15effects,amin15evaluation},
but the authors note that this could be due to the simple 3D stimuli used in the test.

Background EEG readings can also serve as a measure of fatigue \cite{li08measurement}.
Chen et al.\ found that the gravity frequency of the EEG power spectrum and power spectral entropy 
decrease after prolonged periods of watching 3D TV and showed that these measurements can 
act as a predictor of fatigue \cite{chen13eeg}.
Both gravity frequency and power spectral entropy are
decreased greatly on frontal and temporal, and especially
in the prefrontal region after continued 3D viewing, while the effect on gravity was 
not significant on parietal and central areas
\cite{chen14assessment}.

Models based on alpha, beta and theta activities have been proposed in the literature.
Power of high-frequency components which are associated with stress, including the beta
band, increases during 3D viewing \cite{li08measurement,kim11eeg}.
Frey et al.\ measured a power decrease in the alpha band and increases in theta and beta bands 
in the parietal area Pz when viewing non-comfortable content \cite{frey14assessing}.
However, beta activity seems to reduce with the onset of fatigue. 
Zou et al.\ found a significant increase in alpha and a reduction in beta activity 
after prolonged stereoscopic viewing and the onset of fatigue, and suggest that alpha may be the most 
promising index for measuring fatigue \cite{zou15eeg}.
Zhao et al.\ developed a multi-variate regression model of fatigue based on a combination
of different frequency bands \cite{zhao11multivariate}. 

\subsubsection{MRI}
It is known that activity in the occipital lobe (notably V3) and parietal lobe (notably MT) is
very sensitive to disparity and correlated with fatigue \cite{backus01human}.
A large fMRI study by Chen et al.\ found that processing information at different depth significantly
affects brain function. After prolonged 3D viewing, there were changes in brain areas BA17, BA18 and 
BA19 (the latter contains V3,V4, and MT), 
which are related to visual search, as well as in the Frontal Eye Field in B8, 
which is associated with uncertainty and expectation \cite{chen15visual}.
Kim et al.\ also found changes in the FEF for stimuli outside of the comfort zone \cite{kim11human}.

A set of experiments by Jung et al.\ examined the brain's response to excessive disparities which
are a common cause of reported discomfort. They found that 
the right middle frontal gyrus (MFG), the right inferior frontal gyrus (IFG), the right intraparietal lobule (IPL), 
the right middle temporal gyrus (MTG), and the bilateral cuneus were significantly activated during the 
processing of excessive disparities, compared to those of small disparities \cite{jung13subjective,jung15towards}.
They conclude that discomfort due to excessive disparities involves both sensory and motor phenomena.
In a comparison, between high-fatigue and low-fatigue groups, the high-fatigue group showed more 
activation at the intraparietal sulcus (IPS) than the low-fatigue group, 
when viewing an excessive disparity stimulus \cite{kim14fmri}, hinting at the increased strain on 
visual attention and eye movement control.


\section{Considerations for emerging technologies}
\label{secHMD}

Increased popularity of head-mounted displays (HMDs) and mobile devices brought some specific
issues related to these devices. Ukai and Howarth note problems with the technology in early HMDs which 
caused fatigue and eye strain \cite{ukai08visual}
Stereoscopic HMDs require a strict alignment of axes, and any small errors will increase symptoms related
to geometric misalignment.
They also tend to increase the feeling of visually-induced motion sickness (VIMS) due to the inconsistency
of visual and other vestibular cues (such as gravity and acceleration) 
\cite{ukai03counter}. This problem
is alleviated the the latest HMDs which incorporate high quality head-tracking, but latency has to be very low. Even so,
there are still conflicts with the vestibular system as a result of constrained peripheral vision \cite{moss11char}.

An early evaluation of HMDs did not find a difference from viewing stereoscopic images on a desktop 
computer \cite{peli98visual}, but more recent studies certainly found important factors. For one, the screen
is extremely close to the eye causing a strong accommodative response, yet the eyes may converge onto a point
in a distance. Hence, nearby displays have been shown to be less comfortable \cite{shiomi13simultaneous}. Since
an HMD must be close to the eye by design, the only viable solution may be to exploit the effect that blur 
can have on accommodation \cite{cho96study}. Early eye-tracking prototypes implementing this on HMDs 
exist, but are still in early stages and do not improve comfort at the moment \cite{duchowski14reducing}. 
In a comprehensive review of user factors affecting HMD users, Patterson et al.\ recommend a fixed vergence
angle, a wide field of view, and a set of recommendation for maximum disparity and angular change for 
successful binocular fusion \cite{patterson06perceptual}.

With the increased popularity of smartphones and tablet computers, multimedia content is increasingly 
viewed ``on-the-go''. This presents several challenges for content creation. First of all, many
discomfort factors are stronger on small displays \cite{cho12measurement}. Secondly, small hand-held devices
can lead to fast motion, which is known to be uncomfortable \cite{lambooij11visual}, especially 
movement in depth \cite{speranza06effect}. Finally, mobile content is frequently created through automated
retargetting methods. Stereo retargeting methods have only recently started taking user comfort
into account \cite{li15onadjustment}. These automated methods are complicated by the fact that mobile 
devices are held at different distances, according to the situation and preference.
Some automated systems have been proposed for hand-held telephony \cite{mangiat12disparity}. 

Since the level of discomfort depends on factors
such as stereoaccuity \cite{kim13influence}, algorithms for discomfort reduction are likely to become 
personalised, and use integrated cameras to incorporate gaze tracking for real-time processing
\cite{bernhard14effects}.

\section{Discussion}
\label{secDiscussion}

There is a large amount of data on discomfort today, and it suggests that visual discomfort
and visual fatigue are very complex phenomena, encompassing many factors. 
Ophthalmological readings confirm a decrease in the eyes' ability to adapt as a result of fatigue.
In the brain, increased activity in the parietal
and occipital lobes indicates higher levels of visual processing, and fatigue has been associated with
relatively low-level processing in the V3 area of the visual cortex, as well as in V4 and MT. But fatigue
seems to also be related to higher cognition in the prefrontal cortex, as well as the areas related to
occular control. The resulting stress can be detected on background EEG readings, and a correlation on
the autonomic nervous system has been established. All this points to large amounts of cognitive stress
on many different parts of the brain.

Consequently, fatigue manifests itself through many symptoms, from dry and sore eyes to headaches, disorientation
and dizziness, hinting
at a wide range of causes including both ocular and neural fatigue. 
To our knowledge, no systematic studies linking specific symptoms to specific 
neurophysiological causes has been performed, but it seems apparent that the effects are interconnected
and that comfortable viewing must aim to eliminate as many causes as possible.

A large body of literature has identified problematic types of content. Misalignment, excessive disparities,
unnatural distortions, fast movement in depth, crosstalk and the overload of the accommodation-vergence
mechanism have consistently resulted in discomfort and fatigue, and much research has gone into reducing these
effects. Fortunately, excessive disparities, distortions, fast movement and crosstalk can be significantly
reduced or eliminated during acquisition, in postprocessing, or through better display technology. Useful 
sets of guidelines have been produced \cite{mendiburu09,zilly11production}, which have proved useful
in reducing fatigue, but have failed to eliminate it completely \cite{zeri15visual}. Fast motion in depth
and excessive disparities have been tackled by computer vision researchers, and post-processing
algorithms for reducing discomfort exist \cite{didyk11perceptual}.

Misalignment, lack of parallax cues, and the accommodation-vergence conflict are more difficult to tackle.
The development of new, mobile technologies and head-mounted displays poses new challenges. Due to the mobility
of hand-held devices, the viewing geometry can vary more than in a classic cinema or TV watching situation,
causing misalignment. Such devices are viewed from close distances, which can lead to additional overload of the
accommodation-vergence mechanism. This is especially true of head-mounted displays, which put the display very 
close to the eye, but simulate a natural environment and distant objects. Real-time systems which react to the
viewer's position and gaze direction exist as prototypes, but represent early stages of 
research \cite{duchowski14reducing,bernhard14effects}. The effect of blur on controlling accommodation 
could play an important role here, but only if it is fast and accurate enough to simulate natural viewing
conditions without causing additional strain.

\section{Conclusion}
\label{secConclusion}

We have presented a comprehensive review of literature regarding stereoscopic viewing discomfort. We have
incorporated knowledge from a variety of disciplines, as well as the latest neurophysiological findings,
in order to present a complete and balanced picture of the state of research on this topic. The data
suggest that there are many causes of discomfort, and that unnatural stereoscopic viewing affects all stages
of visual processing. The wide range of discomfort symptoms is a natural consequence of this.

Solutions to these problems are needed if stereoscopic 3D is to become more popular. They will require
further cooperation across fields, and this cooperation has already begun, with new algorithms from image
processing, new display technologies and new perceptual models for predicting and understanding discomfort.
We hope that this review proves useful to researchers in this field.

\section{Acknowledgements}

This work was supported by UK EPSRC grant EP/M01469X/1.

\bibliographystyle{plain}
\bibliography{terzic_hansard_3d}

\end{document}